\documentclass[twocolumn,preprintnumbers,amsmath,amssymb,floatfix]{revtex4}

\usepackage{graphicx}
\usepackage{dcolumn}
\usepackage{bm}
\usepackage{epsf}
\usepackage{amsmath}
\usepackage{amsfonts}

\newcommand{\DDir}{\relax{D\kern-.7em{/}}}

\newcommand{\haf}{\frac{1}{2}}

\newcommand{\ra}{\rightarrow}

\newcommand{\X}{\times}

\newcommand{\bj}{\textbf{j}}

\newcommand{\br}{\textbf{r}}

\newcommand{\be}{\begin{equation}}
\newcommand{\ee}{\end{equation}}
\newcommand{\bea}{\begin{equation*}}
\newcommand{\eea}{\end{equation*}}

\newcommand{\abs}[1]{\left\vert#1\right\vert}

\newcommand{\ave}[1]{\left\langle #1\right\rangle}

\newcommand{\pr}{\partial}

\newcommand{\nin}{\relax{\in\kern-.8em{/}}}

\newcommand{\lm}{\lambda}

\newcommand{\Om}{\Omega}
\newcommand{\om}{\omega}

\newcommand{\vep}{\varepsilon}
\newcommand{\ep}{\epsilon}

\newcommand{\yr}{\mbox{ yr}}

\newcommand{\bz}{\textbf{z}}
\newcommand{\hz}{\hat\bz}

\newcommand{\by}{\textbf{y}}
\newcommand{\hy}{\hat{\by}}

\newcommand{\bna}{\mathbf{\nabla}}
\newcommand{\bbj}{\textbf{j}}

\newcommand{\bee}{\textbf{e}}
\newcommand{\bxx}{\textbf{x}}
\newcommand{\hx}{\hat \bxx}

\newcommand{\hj}{\hat \bbj}

\newcommand{\Koz}{\text{Koz}}
\newcommand{\Quadratic}{\text{Quadratic}}
\newcommand{\Obl}{\text{Obl}}
\newcommand{\Tru}{Tr(\mathbf{u})}
\begin{document}
\title{Exponential growth of eccentricity in secular theory}
\author{Boaz Katz\footnote{John Bahcall Fellow, Einstein Fellow}}
\author{Subo Dong\footnote{Sagan Fellow}}
\affiliation{Institute for Advanced Study, Princeton, NJ 08540, USA}
\begin{abstract}
The Kozai mechanism for exponentially exciting eccentricity of a Keplerian orbit by a distant perturber is extended to a general perturbing potential. In particular, the case of an axisymmetric potential is solved analytically. The analysis is applied to orbits around an oblate central object with a distant perturber. If the equatorial plane of the central object is aligned with the orbit of the distant perturber (axisymmetric potential), a single instability zone, in which eccentricity grows exponentially, is found between two critical inclinations; if misaligned (non-axisymmetric potential), a rich set of critical inclinations separating stable and unstable zones is obtained\cite{Vashkoviak74}. The analysis is also applied to a general quadratic potential. Similarly, for non-axisymmetric cases, multiple stability and instability zones are obtained. Here eccentricity can reach very high values in the instability zones even when the potential's deviation from axisymmetry is small. 
\end{abstract}
\maketitle
A striking aspect of the secular evolution of a Keplerian orbit weakly perturbed by a distant orbiting mass is that the eccentricity can grow exponentially to high values \cite{Lidov62,Kozai62}. This so-called Kozai mechanism operates in a finite range of mutual inclinations $39.2^o<i<140.8^o$. The high eccentricity excited from initially nearly circular orbits by the Kozai mechanism is suggested to play an important role in the formation and evolution of  many astrophysical systems \cite[e.g.][]{Kozai62,Heisler86,Blaes02,Fabrycky07,Perets09,Thompson10}. 
In this letter we generalize this mechanism by studying the stability of precessing circular orbits for a general perturbing potential. 

Consider a test particle orbiting a central mass $M$ subject to a small, time-independent perturbation by a potential $-\Phi(\br)$, so that the total potential is   
\begin{equation}
V(r)=-\frac{GM}{r}-\Phi(\br).
\end{equation} 

At any given time the orbit is approximately Keplerian, with the orbital parameters changing with time. 
In the secular approximation, the Hamiltonian is averaged over the orbit's phase (mean anomaly) to obtain equations of motion for the orbital parameters. Under this approximation, the semi-major axis $a$ is constant in time. The problem reduces to understanding the long-term evolution of the remaining 4 orbital elements describing an orbit at the given semi-major axis. 

It it useful to describe the orbit of the particle by two dimensionless vectors: $\bj=\mathbf{J}/\sqrt{GMa}$, where $\mathbf{J}$ is the specific angular momentum vector;  $\bee$, a vector pointing in the direction of the pericenter (point of closest approach) with a magnitude equal to the eccentricity $e$. Note that $\bee$ and $\bj$ satisfy 
\begin{equation}\label{eq:ejPhysical}
j^2=|\bj|^2=1-e^2,~~\bee\cdot \bj=0,
\end{equation}
leaving 4 independent parameters. 
      
The equations of motions for these variables are \cite{Milankovich39,Allan63,Tremaine09}
\begin{equation}\label{eq:NLEOM}
\frac{d\bj}{d\tau}=\bj\X\nabla_{\bj}\phi+\bee\X\nabla_{\bee}\phi,~~~\frac{d\bee}{d\tau}=\bj\X\nabla_{\bee}\phi+\bee\X\nabla_{\bj}\phi,
\end{equation}
where
\begin{equation}
\phi(\bj,\bee)=\frac{\ave{\Phi}}{\Phi_0},~\tau=t/t_{\rm sec},~t_{\rm sec}=\frac{\sqrt{GMa}}{\Phi_0},
\end{equation}
and $\ave{\Phi}(\bj,\bee)$ is the potential time-averaged over an orbit set by $\bee$, $\bj$ and the fixed $a$, which is conserved in time.  
When using equations \eqref{eq:NLEOM}, the 6 components of  $\bee$ and $\bj$ should be considered independent variables, while only the solutions that satisfy the physical conditions Eq. \eqref{eq:ejPhysical} should be considered.  Note that these physical conditions represent a gauge freedom that does not affect the equations of motion \cite{Tremaine09}.

Throughout this letter the following examples are considered. 

1. ``Kozai'' - The perturbing potential is produced by a distant orbiting mass $M_{\rm per}$ at semimajor axis $a_{\rm per}\gg a$. The leading term in the expansion in powers of $a/a_{\rm per}$ (quadrupole=$(a/a_{\rm per})^2$) of the potential averaged in time over the perturber's orbit, is given by  
$\Phi_{\Koz}(\br) =GM_{\rm per}(4 b_{\rm per}^3)^{-1}[r^2-3z^2]$ 
where $b_{\rm per}=a_{\rm per}(1-e_{\rm per}^2)^{1/2}$ is the semi-minor axis of the perturber's orbit and $\hz$ is the direction of the perturber angular momentum. By choosing $\Phi_{0,\Koz} = 3GM_{\rm per}a^2/(8{b_{\rm per}}^3)$, the resulting potential in terms of $\bee$ and $\bj$ is \cite{Tremaine09}
\begin{equation}
\phi_{\Koz} = j_z^2 - 5e_z^2 + 2e^2 -\frac13.\label{eq:phi_Koz}
\end{equation}

2. ``Oblateness'' -  The perturbing potential is the quadrupole potential arised from the oblate central body and is given by  ${\Phi_{\Obl}(\br) = GMJ_2R^2(2r^5)^{-1}[r^2 - 3z^2]}$, where $J_2$ is the gravitational quadrupole  coefficient and $\hz$ is the direction of the central body's spin axis. By choosing 
${\Phi_{0,\Obl} =3J_2 GM R^2/(16 a^3)}$, the resulting potential is \cite{Tremaine09}
\begin{align}
&\phi_{\Obl} =4 \frac{j_z^2 - \frac13j^2}{j^5},\label{eq:phi_Obl}
\end{align}

3. ``Koz-Obl'' - The perturbing potential is a combination of the above two and is given by
\begin{equation}\label{eq:Combined}
\phi_{\Koz-\Obl} = \phi_{\Koz} + \ep\phi_{\Obl},
\end{equation}
where $\Phi_{0,\Koz-\Obl} = \Phi_{0,\Koz}$ is chosen and ${\ep = \Phi_{0,\Obl}/\Phi_{0,\Koz}}$. 
Note that the $\hz$ axes of $\phi_{\Koz}$ and $\phi_{\Obl}$ are not necessarily the same and the inclination between them is denoted as $i_{\Koz-\Obl}$.  

4. ``Quadratic'' - The perturbing potential is a general quadratic function of 
the spatial coordinates, and by a proper choice of coordinate system, it can be
expressed as  
$\Phi_{\Quadratic}(\br) 
=(2\pi G/3) \rho_{\rm eff} [(1+\delta+\sigma)r^2-3(z^2+\delta y^2)]$.
The parameter $\rho_{\rm eff}$ has dimension of density, and is related
to the local matter density $\rho_{\rm local}$ by the Poisson equation, 
which reads $\rho_{\rm eff} \sigma = \rho_{\rm local}$. Such a quadratic 
potential represents the leading order of the potential that arises from 
any mass distribution that is approximately constant in the vicinity of the 
test particle's orbit. In particular, it reduces to the Kozai potential when 
$\delta = \sigma = 0$. Examples include the Galactic tide  \cite[e.g.][]{Heisler86} and a combination of multiple distant orbiting perturbers. By choosing 
$\Phi_{0,\Quadratic} = \pi G\rho_{\rm eff} a^2$, we get
\begin{equation}\label{eq:Quadratic}
\phi_{\Quadratic} = j_z^2+\delta j_y^2 - 5(e_z^2 + \delta e_y^2) + 2(1+\delta +\sigma/2) e^2.
\end{equation}
\emph{Formulation of the Problem}
The problem under study is to find the conditions under which Eqs. \eqref{eq:NLEOM} have precessing circular orbit solutions that are unstable and lead to exponential growth of the eccentricity. 

For circular orbits to be a solution to Eqs. \eqref{eq:NLEOM}, it is sufficient  that 
$
\bna_{\bee}\phi |_{\bee=0}=0. 
$ Henceforth we consider potentials with reflection symmetry $\Phi(\br)=\Phi(-\br)$ for which the leading order in $\bee$ is quadratic and thus they satisfy the above requirement. To the first order in $e$, equations Eq. \eqref{eq:NLEOM} can then be written as,
\begin{align}
&\frac{d\bj}{d\tau}=\bj\X\nabla_{\bj}\phi,\label{eq:LEOMj}\\ 
&\frac{d\bee}{d\tau} = \mathbf{M}(\bj)\bee,\label{eq:LEOMe}
\end{align}
where the coefficients of the matrix $\mathbf{M}(\bj)$ defined according to ${de_{k}/d\tau=M_{kn}e_n}$ are given by,
${M_{kn}=\vep_{klm}j_l \pr_{e_m,e_n}\phi|_{\bee=0}+\vep_{knm}\pr_{j_m} \phi|_{\bee=0}}$.

Circular orbits precess periodically according to Eq. $\eqref{eq:LEOMj}$.  
To see that the precession is periodic, note that $\bj$ is confined to the sphere $|\bj|=1$ and follows the closed contour lines of $\phi(\bj,\bee=0)=$const. 

To first order in $e$, the trajectory of $\bj(t)$ is the same as that of the circular orbit. 
The problem reduces to solving the linear equation Eq. \eqref{eq:LEOMe}, with time-varying, periodic coefficients according to the periodic trajectory of $\bj(t)$.

\emph{Axisymmetric potentials}
Consider first axisymmetric potentials (with reflection symmetry). These include the Kozai, oblateness, quadratic (with $\delta = 0$) potentials introduced above as well as other potentials such as the averaged potential of a perturber on a circular orbit to all orders in the multipole expansion \cite{Kozai62} or the potential of a proto-planetary disk \cite{Terquem10}. By eliminating $j$ using the gauge freedom $j^2 = 1- e^2$, such potentials can be written in the following form 
\begin{equation}\label{eq:AxisymmetricPot}
\phi = a_0 + \haf a_I e^2 + \haf a_z {e_z}^2 + O(e^4),
\end{equation}
where $a_0$, $a_I$ and $a_z$ are functions of $j_z$. Note that ${\nabla_{\bee}\phi=a_I\bee+a_ze_z\hat z}$. 
In this case the unperturbed orbit's angular momentum precesses around $\hat z$ according to $d\bj/d\tau=\dot\Om\hat z\times \bj$ (Eq.\eqref{eq:LEOMj}), where $\dot\Om=d\Om/d\tau=-\pr_{j_z}\phi$ is the constant precession frequency and $\Om$ is the longitude of the ascending node (angle between $\hz\times \bj$ and $\hat x$).  It is useful to work in a rotating reference frame, $\hz'=\bj$, $\hx'=\hz\times\hj/\sin i $ and $\hy'=\hz'\times\hx'$ precessing with $\bj$, where $i$ is the orbit's inclination (angle between $\bj$ and $\hz$) which is constant in the linear approximation. Note that $e_{x'}=e\cos\om,e_{y'}=e\sin\om$ where $\om$ is the argument of pericenter (angle between $\bee$ and $\hz\times \bj$). 

The linear equation, Eq.\eqref{eq:LEOMe}, written in this frame, and restricted to physical eccentricity vectors perpendicular to $\bj$ (vectors satisfying $e_{z'}=0$) is 
\begin{equation}\label{eq:LEOMaxi}
\frac{d}{d\tau}\left(\begin{array}{c}e_{x'}\\e_{y'}\end{array}\right)=\left[\begin{array}{cc}
0&-a_I-a_z\sin^2i\\
a_I&0\\
\end{array}\right]\left(\begin{array}{c}e_{x'}\\e_{y'}\end{array}\right).
\end{equation}
In Eq. \eqref{eq:LEOMaxi}, the coefficients $a_I$ and $a_z$ are to be evaluated at $e=0$ and thus are functions of $i$ alone and are time independent. 
The eigenvalues are $\lm_{\pm} = \pm\sqrt{\Delta}$, where
\begin{equation}\label{eq:lam}
\Delta=-a_I(a_I+a_z\sin^2i)
\end{equation}
and eccentricity grows exponentially if and only if  
$
\Delta>0.
$
For the cases of Kozai, oblateness and quadratic with $\delta=0$ (Eqs. \eqref{eq:phi_Koz}, \eqref{eq:phi_Obl} and \eqref{eq:Quadratic}),  
$a_{I,\Koz} = 4$ and $a_{z,\Koz} = -10$, 
$a_{I,\Obl} = 20\cos^2 i-4$ and $a_{z,\Obl}  = 0$, and $a_{I,\Quadratic} = 4 + 2\sigma$ and $a_{z,\Quadratic} = -10$, respectively. The instability criterion for the Kozai Mechanism, $i>i_c=\sin^{-1}(\sqrt{2/5})$ \cite{Lidov62,Kozai62}, is reproduced while circular orbits are stable for all inclinations in the oblateness case.

Consider next a test particle perturbed by the combined effects of oblateness and Kozai (Eq. \eqref{eq:Combined}) in the case where they are aligned (i.e. $i_{\Koz-\Obl}=0$). One example is a satellite of Jupiter perturbed by its oblateness and the Sun (the spin axis of Jupiter is aligned with its orbital angular momentum to about $3^o$). In this case $\phi$ has coefficients $a_I=4[1+\ep(5\cos^2 i-1)]$ and $a_z=-10$.  

Using the instability condition $\Delta>0$,  for $\epsilon\leq1 (\epsilon>1)$ instability occurs at inclinations $i>i_{c,1}$ (in the range $i_{c,1}<i<i_{c,2}$) where \cite{Vashkoviak74}
\begin{align}\label{eq:critIncizin0}
&\sin^2i_{c,1}=\frac45 \left(1-\frac{1}{4\epsilon+2}\right),\sin^2i_{c,2}=\frac45\left(1+\frac{1}{4\epsilon}\right).
\end{align}
In the limit $\epsilon\ra 0~(\infty)$, the instability zone of Kozai (oblateness) is reproduced. Note that for large $\ep$ the unstable zone reduces to a small interval with a width $\Delta i\approx 2/(5\epsilon)$ in the vicinity of the so-called ``critical inclination''  $i_c=\sin^{-1}((4/5)^{1/2})$ \cite[see e.g.][for discussion of the critical inclination]{Jupp88}. 

\emph{General perturbing potentials}
Consider next a general perturbing potential. The analysis can be proceeded similarly to the axisymmetric case by working in the above-mentioned rotating frame, where in general $i$ is not constant but varies periodically with time. The equations of motion \eqref{eq:LEOMe} have the form
$
d\bee'/d\tau=\mathbf{m}(t)\bee'
$
where $\mathbf{m}$ is a $2\X2$ matrix with coefficients which change with time (periodically) and therefore cannot be solved analytically in general.  

The stability of the solution $\bee'=0$ (precessing circular orbit) can be studied by studying the linear operator $\mathbf{u}$ which represents the integration of the above equation over one period $T$ and is defined by $\bee'(T)=\mathbf{u}\bee'(0)$ \cite[Floquet anlysis, e.g.][section 28]{Arnold73}. After $n$ periods $(t = nT)$, $\bee'(nT) = \mathbf{u}^n\bee'(0)$. The condition for instability is that $\mathbf{u}$
has an eigenvalue $x$ with magnitude larger than unity. The corresponding eigenvector $\mathbf{e}_I$ satisfies 
$\mathbf{u}^n \bee_I = x^n\bee_I$. Given that $\abs{x}>1$, any initial condition 
results in exponential growth of $e$ with time (except possibly for initial $\bee$ which is perpendicular to $\bee_I$). 

Given that $\bee'(0) = \bee(0)$ and $\bee'(T) = \bee(T)$ (since the rotating frame returns to itself after one period), the linear operator $\mathbf{u}$ can be calculated using Eq. \eqref{eq:LEOMe} directly, without moving to the rotating frame. This can be done by choosing any two physical initial conditions $\bee_l(0)$, ($l=1,2$),integrating them using Eq. \eqref{eq:LEOMe}, and finding $\mathbf{u}$ by requiring $\bee_{l}(T)=\mathbf{u}\bee_{l}(0)$.

The analysis of the eigenvalues of $\mathbf{u}$ is simplified by the fact that for the equations considered $\text{det}(\mathbf{u})=1$. This can be easily shown by relating $\mathbf{u}$ to the $3\times3$ operator $\mathbf{U}$ representing the integration of eq. \eqref{eq:LEOMe} in the inertial frame. Since $Tr{(\mathbf{M})}=0$ (can be seen directly form the expression of the coefficients of $\mathbf{M}$ following \eqref{eq:LEOMe}), we have $\text{det}(\mathbf{U}) = 1$ which leads to $\text{det}(\mathbf{u}) = 1$.


The eigenvalue equation for $\mathbf{u}$ is $x^2 - Tr(\mathbf{u})x + 1 = 0$ and the problem reduces to finding the value of $Tr(\mathbf{u})$.  For $|Tr{(\mathbf{u})}| > 2$, there is a real eigenvalue $x> 1$ and the orbit is unstable. If $|Tr{(\mathbf{u})}| \leq 2$, both eigenvalues have magnitude $|x| = 1$ and the orbit is stable.

For axisymmetric potentials, one cycle takes $T=2\pi/\dot\Om$ implying that
\begin{equation}\label{eq:TruKozObl}
Tr(\mathbf{u})=e^{\lm_{+} T}+e^{\lm_{-} T}=2\cosh\left(2\pi\sqrt{\Delta}/\dot\Om\right),
\end{equation}
where $\Delta$ is given by Eq. \eqref{eq:lam}. The condition $\abs{Tr(\mathbf{u})>2}$ is satisfied if and only if $\Delta > 0$, in agreement with the analysis above. 

\begin{figure}
\includegraphics[scale=0.7]{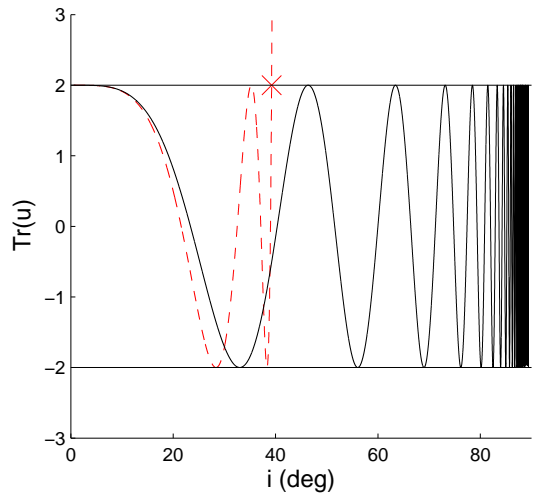}
\caption{\label{fig:kozobl}The value of $Tr(\mathbf{u})$ as a function of inclination, given by Eq. \eqref{eq:TruKozObl} for the axisymmetric Kozai and Oblateness potentials. Instability corresponds to $\abs{Tr(\mathbf{u})}>2$. The red cross corresponds to the Kozai critical inclination.}
\end{figure}

The resulting values of $Tr(\mathbf{u})$ as a function of inclination for the cases of Kozai and  oblateness are shown in Fig. \ref{fig:kozobl}. As can be seen in the figure, while the Kozai case is unstable only above the critical angle, and the oblateness case is always stable, both potentials exhibit a rich structure with many points satisfying $|\Tru|=2$. For a general axisymmetric potential, $|\Tru|=2$ occurs whenever $\Delta \leq 0$ and $\sqrt{-\Delta}/\dot\Om$ is an integer (see Eq. \eqref{eq:TruKozObl}). As shown below, these points can become unstable once small non-axisymmetric potentials are added.

\begin{figure}
\includegraphics[scale=0.7]{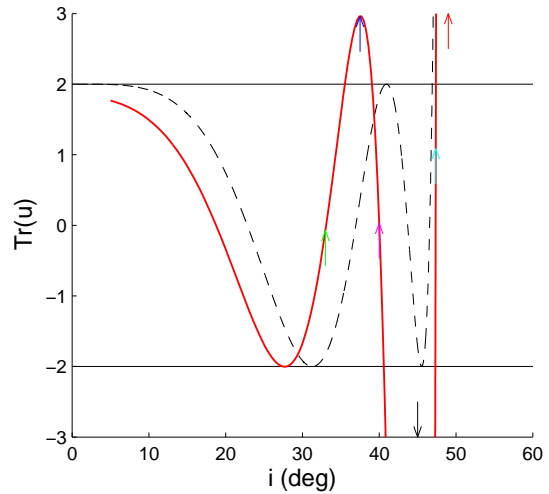}
\caption{\label{fig:tru_ep1}The value of $\Tru$ as a function of inclination for combined Kozai and oblateness potential \eqref{eq:Combined} with $\ep=0.25$ for  $i_{\Koz-\Obl}=0$ (dashed black line) and $i_{\Koz-\Obl}=23^o$ (red solid line). Arrows indicate inclinations used in Fig. \ref{fig:e_of_t_ep1_23deg}.}
\end{figure}

Consider next the combined Kozai and oblateness potential. One example is a satellite of the Earth perturbed by its oblateness and the Moon. For an Earth orbiting satellite perturbed by the Earth's oblateness and the Moon, $\ep$ is related to the
semi-major axis by $a \approx 6.26R_{\rm Earth}\ep^{-1/5}$ with the secular time scale given by $t_{\rm sec}\approx77\ep^{3/10}\yr$.  
The  value of $\Tru$ for the case of $\ep=0.25$ is presented in Fig. \ref{fig:tru_ep1}, for $i_{\Koz-\Obl}=0$ (dashed black line) and $i_{\Koz-\Obl}=23^o$ (red solid line, $23^o$ is the angle between Earth's equatorial plane and the ecliptic plane.). As can be seen the aligned case has one unstable zone above the critical inclination corresponding to Eq. \eqref{eq:critIncizin0} (for $\ep<1$) while the misaligned case has additional unstable zones roughly corresponding to the inclinations in which $|\Tru|=2$ in the axisymmetric case.  
\begin{figure}
\includegraphics[scale=0.7]{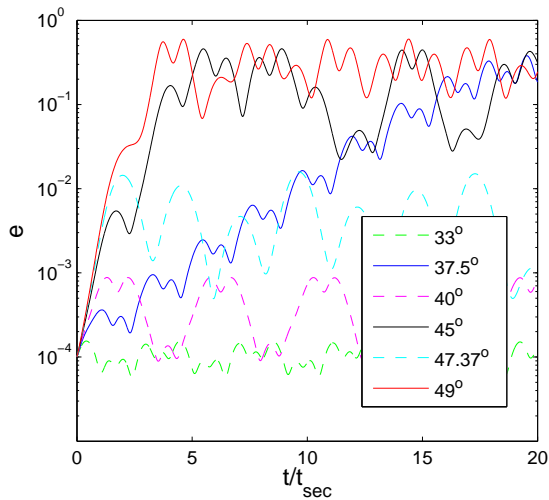}
\caption{\label{fig:e_of_t_ep1_23deg} Evolution of the eccentricity for the combined Kozai (dashed red line) and Oblateness (solid black line) potential with $i_{\Koz-\Obl}=23^o$, for a set of initial inclinations plotted in different colors, as indicated by arrows in Fig. \ref{fig:tru_ep1}.} 
Unstable (stable) orbits are shown in solid (dashed) lines.
\end{figure}
\begin{figure}
\includegraphics[scale=0.7]{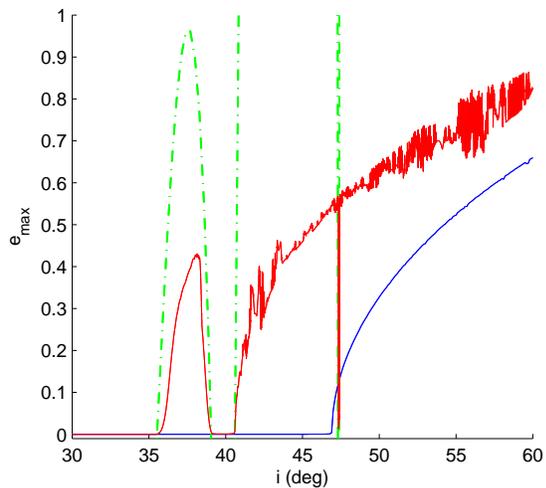}
\caption{\label{fig:emax_koz_obl_ep1}The maximum eccentricity as a function of initial inclination for the combined Kozai and Oblateness potential with $i_{\Koz-\Obl}=23^o$ (red solid). The value of $|\Tru|-2$ from linear analysis  (dashed-dotted green) and the maximum eccentricity obtained in the same time for the aligned combined potential ($i_{\Koz-\Obl}=0$, blue solid) are shown for comparison.}
\end{figure}

The results of numerical integrations of the full (non linear) equations of motion Eqs. \eqref{eq:NLEOM} for a few initial inclinations indicated by arrows in Fig. \ref{fig:tru_ep1} (with initial values $e=10^{-4}, \Om=-\pi/2, \om=0.5$) are presented in Fig. \ref{fig:e_of_t_ep1_23deg}. Integrations with initial inclinations in unstable (stable) zones are shown in solid (dashed) lines.
As can be seen, in the unstable zones the eccentricity grows exponentially to significant values. 

The maximum eccentricity reached in the interval ${t=0-20t_{\rm sec}}$ as a function of initial inclination (with the same initial values $e=10^{-4}, \Om=-\pi/2, \om=0.5$) is shown in Fig. \ref{fig:emax_koz_obl_ep1} (solid red). 
For comparison, the value of $|\Tru|-2$ is shown on the same figure (dashed-dotted green). As can be seen, the linear instability condition $|\Tru|-2>0$ accurately captures the regions where $e$ reaches significant values.

\begin{figure}[h]
\includegraphics[scale=0.7]{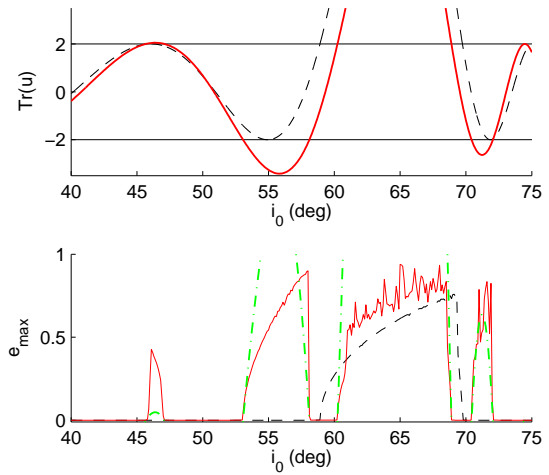}
\caption{\label{fig:multiplot_ep10}Results for combined potential with $\ep=2.5$. The upper panel shows the value $\Tru$ for aligned (black dashed) and $i_{\Koz-\Obl}=23^o$ (red solid). The bottom panel shows the corresponding maximum eccentricity as a function of inclination. $|\Tru|-2$ for $i_{\Koz-\Obl}=23^o$ is shown for comparison (green dashed).}
\end{figure}

Similar results for the combined potential with the same alignments $i_{\Koz-\Obl}=23^o$ and $\ep=2.5$  are presented in Fig. \ref{fig:multiplot_ep10}. Similarly, a rich structure of stable and unstable zones is obtained. 

\begin{figure}[h]
\includegraphics[scale=0.7]{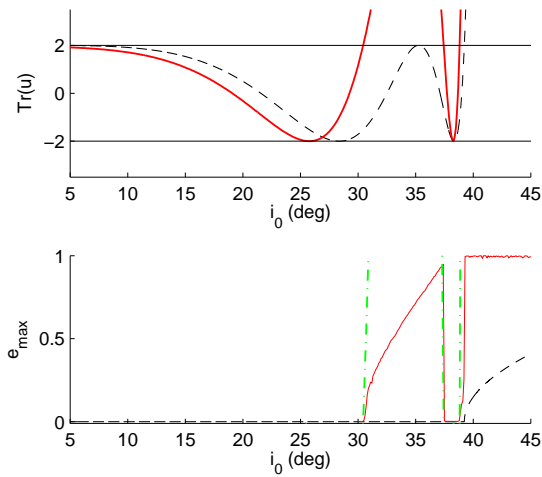}
\caption{\label{fig:quadratic} Results for the quadratic potential with $\sigma = 0$ and $\delta = 0.1$. $\Tru$ as a function of initial inclination is shown in red solid line in the upper panel. For comparison, the case $\sigma = \delta = 0$ (Kozai) is plotted in black dashed. The bottom panel shows the corresponding maximum eccentricity in the interval ${t=0-40t_{\rm sec}}$.  $|\Tru|-2$ for $\delta = 0.1$ is shown for comparison (green dashed). }
\end{figure}

Finally, we perform stability analysis on the quadratic potential 
given in Eq.\eqref{eq:Quadratic} with $\sigma=0$ and $\delta=0.1$,  
which corresponds to a small non-axissymmetric component added to the Kozai
potential (that is, $\sigma=\delta=0$ in Eq. \eqref{eq:Quadratic}). The results 
are shown in Fig. \ref{fig:quadratic} . As can be seen in the upper panel, due to the non-axissymmetry, a new instability zone is formed near $i_0 \sim 35 \deg$, where $\Tru = 2$ for Kozai. As illustrated in the red solid line in the bottom panel, 
within $40 t_{\rm sec}$, substantial eccentricities are obtained in this new zone. Interestingly, the attainable maximum eccentricities beyond the Kozai critical angle are significantly larger than those in the Kozai case (black dashed). In fact, 
upon closer inspection, we find maximum eccentricities reach as high as 
$\sim 0.999$ during short episodes in which the inclinations cross 90 $\deg$, similar to the behaviors obtained when the octupole contribution is added to the 
Kozai quadrupole potential \cite{Ford00, Naoz11}. Detailed discussions are deferred to a future publication.

Upon completion of this work, we learned that the stability analysis on the combined potentials of Kozai and oblateness was performed in \cite{Vashkoviak74}.

We thank Scott Tremaine for providing substantial help. We are grateful to Rashid Sunyaev for pointing out the work by M. Vashkoviak and Renu Malhotra for helpful discussions. B.K. is supported by NASA through Einstein Postdoctoral Fellowship awarded by the Chandra X-ray Center, which is operated by the Smithsonian Astrophysical Observatory for NASA under contract NAS8-03060. Work by SD was performed under contract with the California Institute of Technology (Caltech) funded by
NASA through the Sagan Fellowship Program.

\end{document}